\documentclass[twocolumn]{aastex631}
\usepackage{CJK}
\usepackage{array} 
\usepackage{threeparttable}
\usepackage{amsmath}

\newcommand{\kms}{km $s^{-1}$}

\received{XXX}
\revised{XXX}
\accepted{XXX}

\shorttitle{HVS}
\shortauthors{Xu et al.}

\begin{document}
\begin{CJK*}{UTF8}{gbsn}


\title{Photometric Distances for Metal-poor Giants and a Search for Hypervelocity Stars with LAMOST DR13 and Gaia DR3}

\author[0000-0003-3535-504X]{Shuai Xu (徐帅)}
\affiliation{Institute for Frontiers in Astronomy and Astrophysics, Beijing Normal University,  Beijing 102206, China}
\affiliation{School of Physics and Astronomy, Beijing Normal University, Beijing 100875, People’s Republic of China}

\author[0000-0003-2471-2363]{Haibo Yuan (苑海波)}
\affiliation{Institute for Frontiers in Astronomy and Astrophysics, Beijing Normal University,  Beijing 102206, China}
\affiliation{School of Physics and Astronomy, Beijing Normal University, Beijing 100875, People’s Republic of China}

\author[0000-0002-1259-0517]{Bowen Huang (黄博闻)}
\affiliation{Institute for Frontiers in Astronomy and Astrophysics, Beijing Normal University,  Beijing 102206, China}
\affiliation{School of Physics and Astronomy, Beijing Normal University, Beijing 100875, People’s Republic of China}

\author[0000-0001-8424-1079]{Kai Xiao (肖凯)}
\affiliation{School of Astronomy and Space Science, University of Chinese Academy of Sciences, Beijing, 100049, China}

\author[0000-0002-3956-8061]{Wuming Yang (杨伍明)}
\affiliation{School of Physics and Astronomy, Beijing Normal University, Beijing 100875, People’s Republic of China}

\correspondingauthor{Haibo Yuan}
\email{yuanhb@bnu.edu.cn}
\correspondingauthor{Wuming Yang}
\email{yangwuming@bnu.edu.cn}

\begin{abstract}

Hypervelocity stars (HVSs) are stars with velocities high enough to escape the Milky Way, but their identification depends sensitively on distance estimates, particularly for distant giants. In this work, we search for metal-poor HVS candidates by combining LAMOST DR13 spectroscopy with Gaia DR3 astrometry. We calibrate a metallicity-dependent color--absolute-magnitude relation for normal metal-poor giants using a high-quality reference sample, and apply it to derive photometric distances for 41{,}331 stars. The relation reproduces the reference absolute magnitudes with a scatter of 0.24\,mag, corresponding to an intrinsic distance uncertainty of $\sim$9.7\%. Combining these distances with Gaia proper motions and LAMOST radial velocities, we identify 13 initially unbound candidates under the Galactic potential from \citet{McMillan2017}. Spectral inspection indicates that several are chromospherically active binaries or other non-standard systems for which a giant-star calibration is unreliable; removing these contaminants leaves nine metal-poor HVS candidates. Backward orbit integrations suggest that one candidate is most consistent with a disk origin, 
while three have trajectories suggestive of an association with the Sagittarius stream.

\end{abstract}

\keywords{Hypervelocity stars; Distance measure; Milky Way dynamics}

\section{Introduction} \label{sec:intro} 

High-velocity stars are among the most extreme stellar populations in the Milky Way. A particularly prominent subset are hypervelocity stars (HVSs), which are commonly interpreted as stars ejected from the Galactic Center. The canonical formation channel is the tidal disruption of stellar binaries by the central massive black hole (\citealt{hills1988}; see also \citealt{yu2003}). Since their prediction and first observational discovery (\citealt{brown2005}), HVSs have been used as probes of the Galactic potential, the environment of Sgr~A*, and the dynamical processes operating in the inner Milky Way (\citealt{brown2015}).

Not all high-velocity stars originate in the Galactic Center. Extreme velocities can also be produced by supernova explosions in close binaries (\citealt{blaauw1961,portegies2000,geier2015,gao2019}), dynamical ejections in dense stellar systems (\citealt{leonard1990,gvaramadze2009,gvaramadze2011,schoettler2020,huang2025}), or tidal debris and ejections associated with satellite galaxies (\citealt{abadi2009,boubert2017,erkal2019,huang2021}).

Early searches for HVSs relied primarily on extreme radial velocities of distant early-type stars (\citealt{brown2006,brown2007,brown2014}), establishing the first secure sample of unbound objects. The HVS Survey ultimately identified 21 unbound late B-type stars, implying an all-sky population of order $\sim 300$ unbound 2.5--4\,$M_\odot$ HVSs within 100\,kpc (\citealt{brown2015}).

Large spectroscopic programs---including the Sloan Extension for Galactic Understanding and Exploration \citep[SEGUE;][]{yanny2009}, SEGUE-2 \citep{Rockosi2022}, and the Large Sky Area Multi-Object Fiber Spectroscopic Telescope \citep[LAMOST;][]{cui2012,deng2012,Zhao2012,liu2014}---both added several early-type HVSs (\citealt{zheng2014,huang2017}) and extended searches to lower-mass and later-type stars (\citealt{palladino2014,li2015}). However, in the absence of precise astrometry and reliable distances, the nature of many candidates remained uncertain.

Gaia transformed the field by enabling large-scale searches based on astrometry and, where available, full six-dimensional phase-space information. During the Gaia DR2 era, many studies reported substantial numbers of HVS candidates (\citealt{boubert2018,shen2018,hattori2018,Marchetti2019,du2019,li2020}). Subsequent analyses, however, showed that the robust unbound sample is much smaller: \citet{boubert2018} argued that nearly all previously proposed late-type HVS candidates are likely bound, while \citet{Marchetti2019} and \citet{marchetti2021} found that the number of clean candidates with an escape probability greater than $50\%$ dropped from 125 (DR2) to 94 (EDR3), and to 12 after applying the parallax zero-point correction. This conclusion has been reinforced by later Gaia-based studies, which continued to identify candidates but emphasized contamination, data-quality limitations, and the difficulty of assembling a small, robust sample of genuinely unbound stars (\citealt{Marchetti2022,igoshev2023,li2023,schloz2024,luna2024,Carballo2025}).

Accurate distances are essential for HVS candidate studies because the inferred tangential velocity scales linearly with distance, directly impacting the total velocity, the bound/unbound classification, and any reconstructed origin. In the Gaia era, distance uncertainties have therefore become a key limitation: parallaxes with modest signal-to-noise do not yield reliable distances via simple inversion (\citealt{bailer2015,luri2018}). Bayesian distance inferences are often more robust, but for distant sources with small parallaxes they can become strongly prior-dependent, leaving distance estimation as a dominant contributor to the overall uncertainty budget.

In recent years, photometric and spectrophotometric methods have become practical alternatives for distance estimation in large stellar samples, particularly at large distances. For HVS searches, robust distances to giant stars are especially valuable because giants are intrinsically luminous and can be observed far into the halo, where unbound candidates are most readily found. For distant giants, \citet{xue2014} presented a Bayesian method for SEGUE K giants based on metallicity-dependent RGB calibrations, and \citet{zhang2023dist} applied a similar framework to construct distance estimates for 19,544 LAMOST DR8 K giants in the halo. \citet{huang2022,huang2023} estimated distances and other stellar parameters for tens of millions of stars by combining photometry with Gaia data; for stars lacking reliable parallax-based distances, they derived photometric distances from metallicity-dependent absolute-magnitude calibrations with separate treatments for dwarfs and giants.  Collectively, these studies demonstrate the viability of photometric distance methods for distant samples, while highlighting key systematics tied to the adopted absolute-magnitude calibration, reddening corrections, and (critically) correct stellar-type classification.

This issue is particularly acute for giant samples. Unresolved binaries, chromospherically active systems such as RS Canum Venaticorum (RS~CVn) stars \citep{hall1976}, and other peculiar evolved objects may deviate from the standard color--absolute-magnitude relation calibrated for normal metal-poor giants. 
Here, we define normal metal-poor giants as metal-poor stars on the red-giant branch (RGB) locus in the HR diagram, with no evidence for strong activity, obvious spectral peculiarity, or significant contamination from unresolved companions.
If treated as ordinary giants, their absolute magnitudes---and thus photometric distances---can be systematically biased, with direct consequences for inferred space velocities and the resulting bound/unbound classification. 
Rigorous stellar-type vetting is therefore essential for obtaining reliable photometric distances to distant giant candidates in HVS searches.

In this work, we search for HVS candidates among metal-poor giants from LAMOST DR13. These luminous stars can be traced to large heliocentric distances, but robust distance estimates require both an appropriate absolute-magnitude calibration and careful stellar-type vetting. 
Combining these distances with the available phase-space information, we assess candidate bound/unbound status under a range of Galactic potential models. Finally, because non-standard stellar types can bias the distance calibration, we inspect candidate spectra and remove likely contaminants before investigating the possible origins of the remaining objects.

This paper is organized as follows. Section~\ref{method} describes the construction of the metallicity-dependent absolute-magnitude relation and the sample selection. Section~\ref{hvs} presents the identification of unbound candidates and their orbital analysis. Section~\ref{summary} summarizes our main conclusions.

\section{Data and Methodology}\label{method}

\subsection{LAMOST and Gaia}

LAMOST is a 4\,m quasi-meridian reflecting Schmidt telescope equipped with 4000 fibers over a 5\,deg field of view. Its 13th data release comprises more than ten million low-resolution spectra ($R\approx 1800$) spanning a wide range of stellar types. 
Stellar atmospheric parameters and radial velocities for millions of stars are provided by the LAMOST Stellar Parameter Pipeline (LASP; \citealt{wu2011,LAMOSTDR1}), with typical precisions of $\sim$0.1\,dex in atmospheric parameters and $\sim$5\,\kms\ in radial velocity for spectra with $S/N_g>20$.

Gaia is an ESA astrometry mission designed to map the Milky Way through precise measurements of positions, parallaxes, proper motions, photometry, and radial velocities. Operating near the Sun--Earth L2 point, Gaia repeatedly scans the full sky and delivers an increasingly detailed multidimensional view of the Galaxy and its stellar populations (\citealt{gaiadr1}). Gaia DR3 provides astrometry and broad-band photometry for about 1.8 billion sources, along with a wide range of additional products including BP/RP spectra, radial velocities, variable-star classifications, non-single-star solutions, and astrophysical parameters (\citealt{gaiadr3}).

Relative to Gaia DR2, Gaia DR3 offers improved astrometric precision and reduced systematics, but non-negligible parallax zero-point effects remain and must be treated carefully. Consequently, DR3-based distance estimates can differ substantially from DR2 for individual stars, particularly at large distances where parallaxes have low significance. For HVS studies, these revisions can translate into meaningful changes in inferred tangential velocities and, therefore, in the dynamical classification of candidates.

\subsection{Distance Determination}

To estimate distances for our HVS search sample, we construct a metallicity-dependent color--absolute-magnitude relation, using the Gaia $(BP-RP)_0$ color and spectroscopic metallicity to infer $M_G$, and then convert $M_G$ to distance via the distance modulus. To mitigate systematic effects, we focus on metal-poor giants: we require $\mathrm{[Fe/H]}<-1$ to reduce contributions from younger populations (whose broader age distribution increases scatter in the color--magnitude relation), and we impose $\log g<3.5$ to select intrinsically luminous giants that can be traced to large heliocentric distances.

In the following subsections, we describe the training-sample selection, the calibration procedure and its performance, and finally the application of the calibrated relation to the full sample to produce the distance catalog used in the subsequent analysis.

\subsubsection{Sample Selection}

We cross-match LAMOST DR13 with Gaia DR3 and the Gaia DR3 distance catalog of \citet{bailer2021}. For stars with multiple LAMOST observations, we retain only the spectrum with the highest $S/N_g$. This yields an initial sample of 51{,}205 unique metal-poor giants with $S/N_g>20$.

To improve the reliability of the astrometry and photometry, we apply several quality cuts. We require ${\tt RUWE}<1.4$ to reduce the impact of unresolved binaries on the color--absolute-magnitude relation, and ${\tt visibility\_periods\_used}\geq 8$ to ensure robust astrometric solutions. We also apply the ${\tt phot\_bp\_rp\_excess\_factor}$ cut adopted by \citet{xu2022} to remove sources with problematic Gaia photometry. 
Reddening corrections use the \citeauthor*{SFD1998} (\citeyear{SFD1998}; hereafter SFD) map together with the temperature- and extinction-dependent reddening coefficients of \citet{zhang2023}.
Finally, we remove stars with ${\tt parallax\_over\_error}>5$ and $M_{G0,\rm BJ}>4$, which are inconsistent with RGB stars; here $M_{G0,\rm BJ}$ is computed from the \citet{bailer2021} distances and the reddening correction described above. After applying these cuts, 42{,}909 sources remain.

From this parent sample, we define a high-quality subsample for training and testing the metallicity-dependent color--absolute-magnitude relation. We require a \citet{bailer2021} quality flag of 10033, a relative distance uncertainty $\frac{(d_{lo}-d_{hi})}{2*d_{\rm med}}<0.1$, and ${\tt parallax\_over\_error}>10$，where $d_{{\rm lo},i}$ and $d_{{\rm hi},i}$ are the 16th and 84th percentile distances from \citet{bailer2021} and $d_{{\rm med},i}$ is the median distance. 
Here `10033` is the Bailer-Jones et al. (2021) five-digit quality flag, indicating that the source lies within the magnitude range used for the prior, that both the geometric and photogeometric distance posteriors pass the unimodality test, and that the photogeometric prior uses smoothing-spline models.
To reduce extinction-related systematics, we further require $E(B-V)_{\rm SFD}<0.05$, and we retain only stars with $M_{G0,\rm BJ}<4$ to exclude objects that are clearly too faint to be RGB stars.

To reduce contamination from horizontal-branch and subgiant stars, we apply empirical cuts in the Hertzsprung--Russell (HR) diagram. Horizontal-branch contaminants are removed with a metallicity-dependent rectangular mask: stars with $-2.0<\mathrm{[Fe/H]}<-1.5$ are excluded if they fall within $0.7<(BP-RP)_0<0.91$ and $-4<M_{G0,\rm BJ}<0.9$, while stars with $-1.5<\mathrm{[Fe/H]}<1$ are excluded if they fall within $0.7<(BP-RP)_0<1.0$ and $-4<M_{G0,\rm BJ}<1.0$. No cut is applied for $\mathrm{[Fe/H]}\leq -2.0$. We further suppress subgiant contamination by retaining only stars with $M_{G0,\rm BJ}<5.1\,(BP-RP)_0-1$.

\begin{figure}[htbp]
    \centering
    \includegraphics[width=1\linewidth]{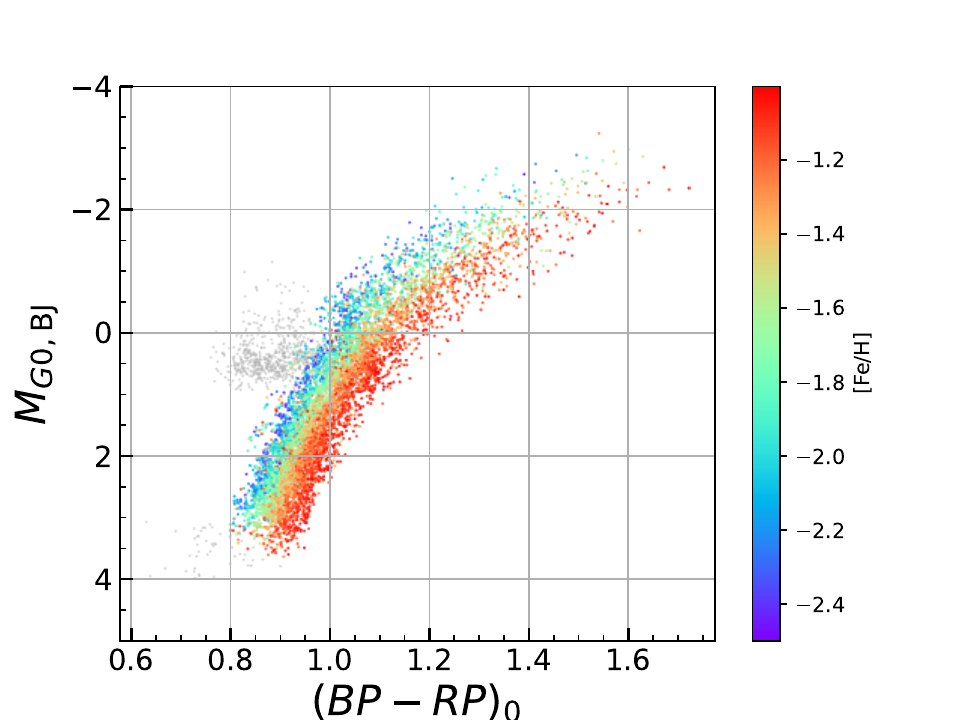}
    \caption{Hertzsprung--Russell diagram of the calibration sample, color-coded by metallicity. Grey dots refer to the dropped horizontal-branch stars and subgiants.}
    \label{fig:hrd_training}
\end{figure}

The resulting calibration sample contains 6298 stars. Its HR diagram is shown in Figure~\ref{fig:hrd_training}, which reveals a clear metallicity dependence in the color--absolute-magnitude relation. We randomly assign 80\% of the sample to the training set and the remaining 20\% to the test set.

\subsubsection{Training and Validation}

We model the metallicity-dependent color--absolute-magnitude relation using polynomial regression with Ridge regularization, adopting $M_{G0,\rm BJ}$ as the reference absolute magnitude for training. 
After an initial fit, we remove $2.5\sigma$ outliers in the residuals and retrain the model on the cleaned sample.
The relation is :
\begin{equation}
\begin{aligned}
M_{G,0} =\;&43.634+14.157\,x-43.257\,y+4.645\,x^2\\
&-10.068\,xy -19.319\,y^2+1.112\,x^3-0.735\,x^2y\\
&+3.715\,xy^2+36.083\,y^3+0.152\,x^4+0.081\,x^3y\\
&+0.029\,x^2y^2 -0.617\,xy^3-11.050\,y^4
\end{aligned}
\end{equation}, 
where $x=[{\rm Fe/H}], y=(BP-RP)_0$.

\begin{figure*}
    \centering
    \includegraphics[width=1\linewidth]{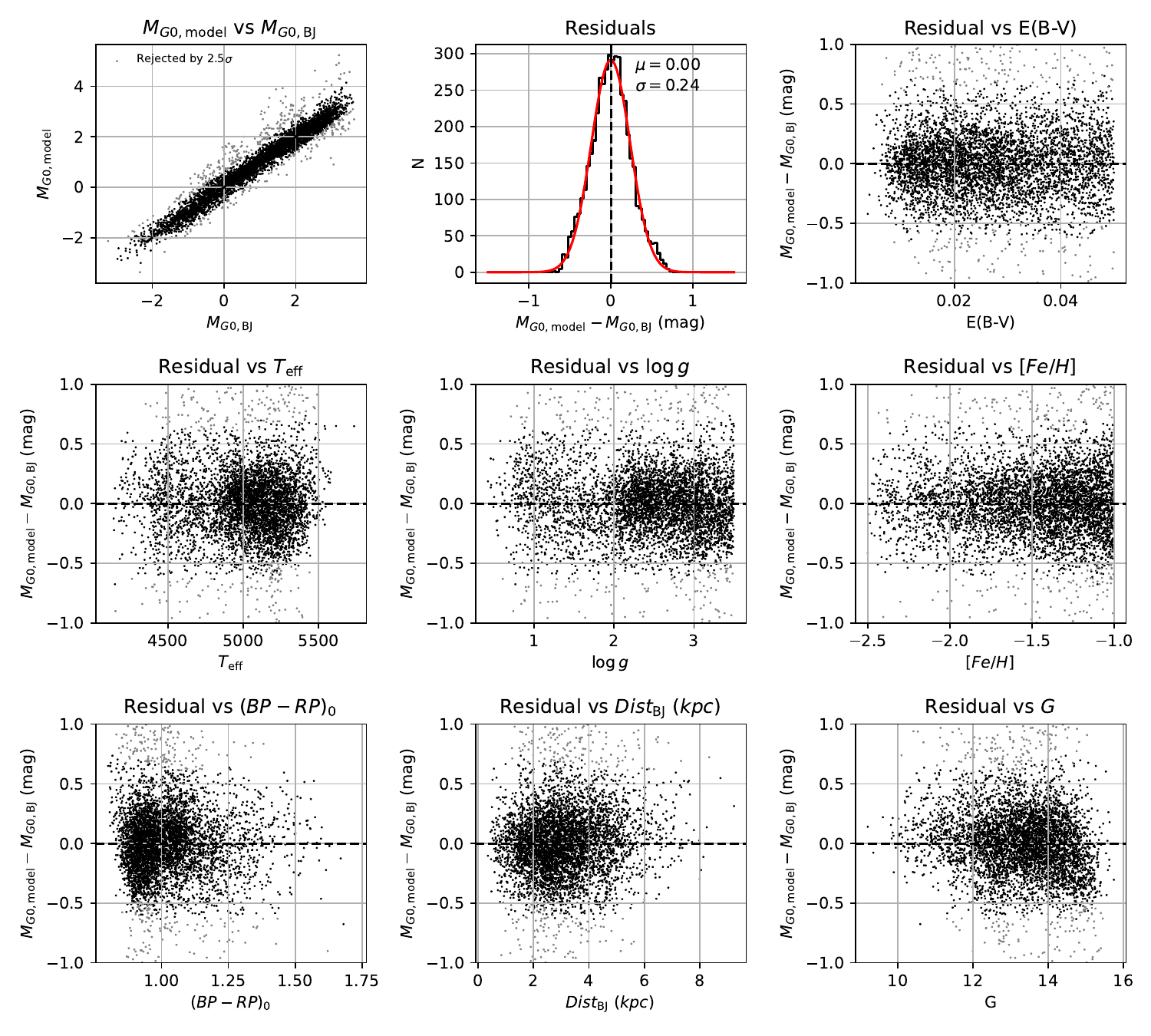}
    \caption{Performance of the metallicity-dependent color--absolute-magnitude relation on the training set. 
    Gray points mark sources rejected by the 2.5\,$\sigma$ clipping.
    Top left: comparison of $M_{G0,\rm model}$ and $M_{G0,\rm BJ}$; the black dashed line indicates the one-to-one relation. 
    Top middle: residual distribution, $M_{G0,\rm model}-M_{G0,\rm BJ}$. The red curve shows a Gaussian fit, with the fitted mean and standard deviation indicated; the vertical dashed line marks zero residual. 
    Remaining panels: residuals versus $E(B-V)_{\rm SFD}$, $T_{\rm eff}$, $\log g$, $\mathrm{[Fe/H]}$, $(BP-RP)_0$, $D_{\rm BJ}$, and $G$. Horizontal dashed lines indicate zero residual.}
    \label{fig:train_data}
\end{figure*}

Figure~\ref{fig:train_data} summarizes the training-set performance. The top-left panel compares $M_{G0,\rm model}$ with $M_{G0,\rm BJ}$, and the top-middle panel shows the residual distribution, defined as $M_{G0,\rm model}-M_{G0,\rm BJ}$. The remaining panels plot residuals versus $E(B-V)_{\rm SFD}$, $T_{\rm eff}$, $\log g$, $\mathrm{[Fe/H]}$, $(BP-RP)_0$, $D_{\rm BJ}$, and $G$. Overall, $M_{G0,\rm model}$ closely reproduces $M_{G0,\rm BJ}$ with no strong systematic trends, and the residuals are approximately Gaussian with $\mu\approx 0$ and $\sigma\approx 0.24$\,mag.

The observed scatter does not reflect the intrinsic model uncertainty alone, because the reference label $M_{G0,\rm BJ}$ inherits random errors from the \citet{bailer2021} distance uncertainties. To estimate the contribution of these label uncertainties to the scatter in $M_{G0,\rm BJ}$, we propagate the distance uncertainty of each training star into absolute magnitude. For the $i$th star, we approximate the 1$\sigma$ distance uncertainty as
\begin{equation}
\sigma_{d,i} = \frac{d_{{\rm hi},i} - d_{{\rm lo},i}}{2}.
\end{equation}
The corresponding absolute-magnitude uncertainty is
\begin{equation}
\sigma_{M_G,i} = \frac{5}{\ln 10}\frac{\sigma_{d,i}}{d_{{\rm med},i}}.
\end{equation}
We summarize the reference-label uncertainty in the training sample by the root-mean-square of these values,
\begin{equation}
\sigma_{\rm BJ,true} = \left( \frac{1}{N}\sum_{i=1}^{N} \sigma_{M_G,i}^2 \right)^{1/2},
\end{equation}
which yields $\sigma_{\rm BJ,true}\approx 0.12$\,mag for our training sample. Assuming the model error and the random error in $M_{G0,\rm BJ}$ are independent, the intrinsic model scatter with respect to the underlying true absolute magnitude is
\begin{equation}
\sigma_{\rm model,true} = \sqrt{\sigma_{\rm model,BJ}^2 - \sigma_{\rm BJ,true}^2}.
\end{equation}
With $\sigma_{\rm model,BJ}=0.24$\,mag and $\sigma_{\rm BJ,true}=0.12$\,mag, we obtain $\sigma_{\rm model,true}\approx 0.21$\,mag, corresponding to a distance uncertainty of $\approx 9.7\%$. Given the stringent distance-quality cuts applied to the training set, this percentile-based propagation should provide a reasonable estimate of the random distance uncertainty.

\begin{figure*}
    \centering
    \includegraphics[width=1\linewidth]{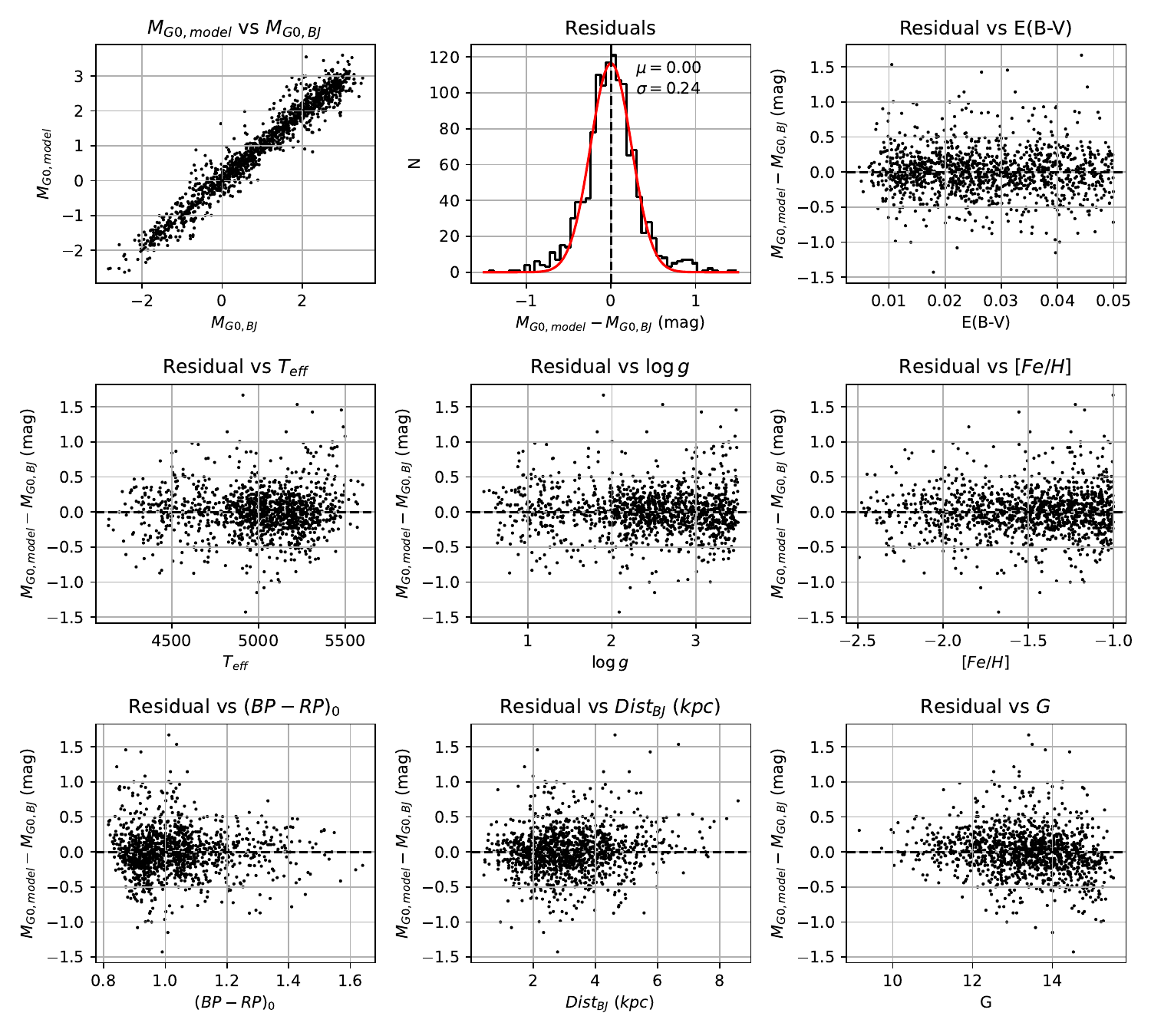}
    \caption{Same as Figure~\ref{fig:train_data}, but for the test set.}
    \label{fig:test_data}
\end{figure*}

We perform the same validation on the test set (Figure~\ref{fig:test_data}). 
The resulting $\sigma$ and $\mu$ are consistent with those of the training set, and the residuals show no obvious systematic trends with any individual observational parameter.

\begin{figure*}
    \centering
    \includegraphics[width=1\linewidth]{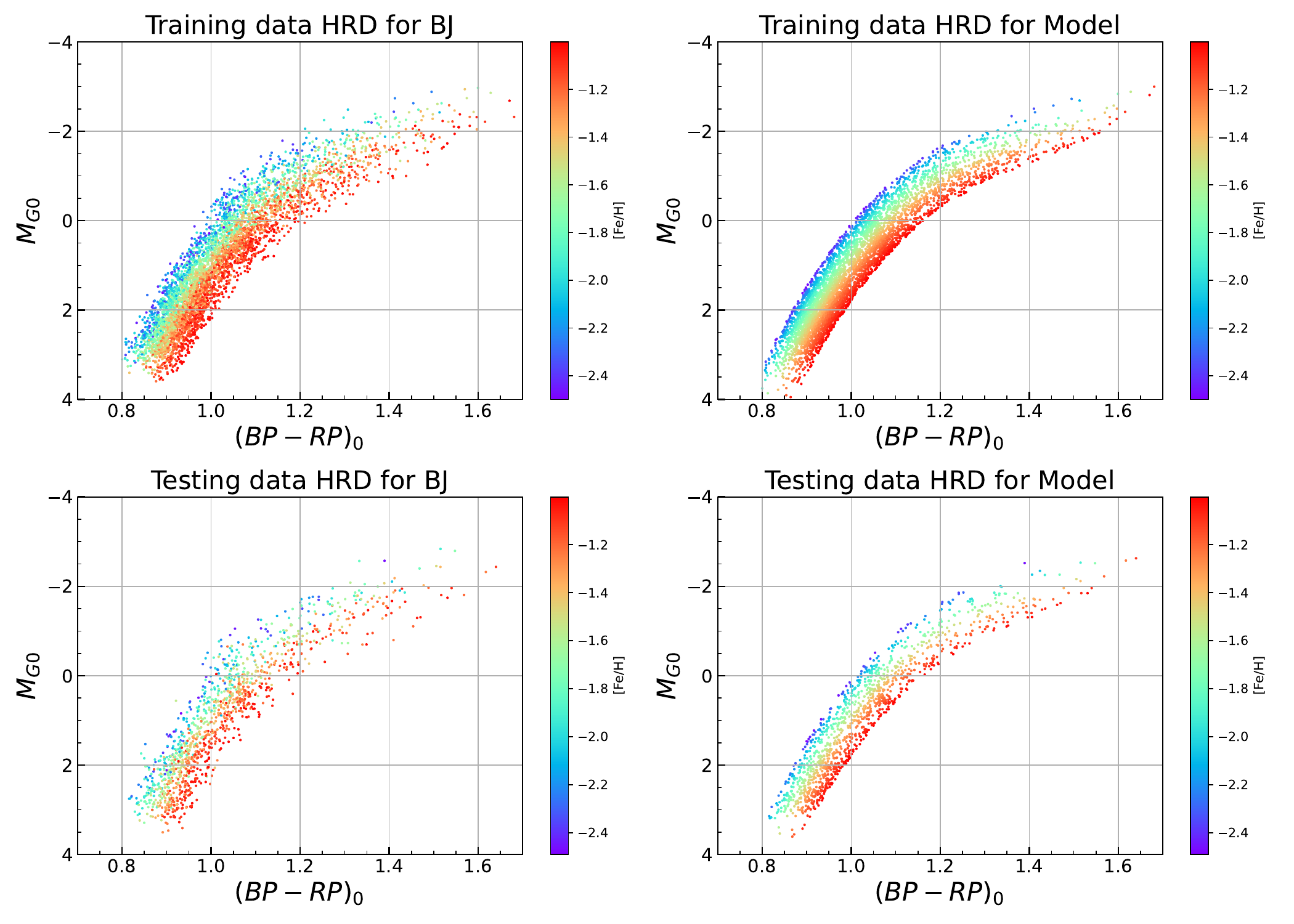}
    \caption{Hertzsprung--Russell diagrams constructed using $M_{G0,\rm BJ}$ (left) and $M_{G0,\rm model}$ (right) for the training sample (top) and test sample (bottom), color-coded by metallicity. The comparison shows how well the model reproduces the RGB morphology and its metallicity dependence.}
    \label{fig:train_test_hrd}
\end{figure*}

In addition to the residual statistics, we assess model performance by visual checking HR diagrams constructed with $M_{G0,\rm BJ}$ and $M_{G0,\rm model}$ for the training and test samples (Figure~\ref{fig:train_test_hrd}). 
Overall, the relation reproduces the main RGB morphology in both samples, but the HR diagrams also highlight its limitations. Toward the relatively metal-rich end of our sample ($\mathrm{[Fe/H]}
\gtrsim -1.3$), the model tends to underestimate $M_{G0}$ for some younger stars, which could bias inferred distances if such objects enter the final HVS sample. The relation also becomes less reliable for redder sources with $(BP-RP)_0>1.6$.

\subsubsection{Distances for the Full Sample}

We apply the calibrated relation to the full parent sample of 42{,}909 sources. 
To stay within the validated domain of the model, we retain only stars with $(BP-RP)_0<1.6$ and $M_{G0,\rm model}<4$, and we convert the predicted absolute magnitudes to distances via the distance modulus. 
The resulting distance catalog contains 41{,}331 stars.
Table \ref{tab:catalog_desc} lists the columns contained in the distance catalog.

\begin{deluxetable*}{lll}
\tablecaption{Description of the machine-readable photometric-distance catalog.\label{tab:catalog_desc}}
\tablewidth{0pt}
\tablehead{
\colhead{Field} & 
\colhead{Description} &
\colhead{Unit}
}
\startdata
\texttt{source\_id} & Gaia DR3 source identifier & -- \\
\texttt{obsid} & LAMOST observation identifier & -- \\
\texttt{ra} & Right ascension from Gaia DR3 & deg \\
\texttt{dec} & Declination from Gaia DR3 & deg \\
\texttt{phot\_g\_mean\_mag} & Gaia DR3 $G$-band magnitude & mag \\
\texttt{phot\_bp\_mean\_mag} & Gaia DR3 $G_{\rm BP}$ magnitude & mag \\
\texttt{phot\_rp\_mean\_mag} & Gaia DR3 $G_{\rm RP}$ magnitude & mag \\
\texttt{ebv\_sfd} & Foreground reddening, $E(B-V)$, from the SFD map & mag \\
\texttt{bp\_rp0} & Extinction-corrected Gaia color, $(G_{\rm BP}-G_{\rm RP})_0$ & mag \\
\texttt{parallax} & Gaia DR3 parallax & mas \\
\texttt{parallax\_error} & Uncertainty in Gaia DR3 parallax & mas \\
\texttt{parallax\_over\_error} & Gaia DR3 parallax signal-to-noise ratio & -- \\
\texttt{pmra} & Proper motion in right ascension, $\mu_{\alpha *}$ & mas yr$^{-1}$ \\
\texttt{pmra\_error} & Uncertainty in $\mu_{\alpha *}$ & mas yr$^{-1}$ \\
\texttt{pmdec} & Proper motion in declination, $\mu_\delta$ & mas yr$^{-1}$ \\
\texttt{pmdec\_error} & Uncertainty in $\mu_\delta$ & mas yr$^{-1}$ \\
\texttt{ruwe} & Gaia DR3 renormalized unit-weight error & -- \\
\texttt{visibility\_periods\_used} & Number of visibility periods used in the Gaia astrometric solution & -- \\
\texttt{phot\_bp\_rp\_excess\_factor} & Gaia BP/RP flux-excess factor & -- \\
\texttt{teff} & LAMOST effective temperature & K \\
\texttt{logg} & LAMOST surface gravity & dex \\
\texttt{feh} & LAMOST metallicity, [Fe/H] & dex \\
\texttt{rv} & LAMOST radial velocity & km s$^{-1}$ \\
\texttt{snrg} & LAMOST $S/N_g$ & -- \\
\texttt{MG0\_pred} & Predicted extinction-corrected absolute magnitude, $M_{G,0}^{\rm pred}$ & mag \\
\texttt{dist\_phot} & Photometric distance inferred from $M_{G,0}^{\rm pred}$ & kpc \\
\texttt{flag} & Source classification flag: RGB-like or SSG-like & -- \\
\enddata
\tablenotetext{}{
The full machine-readable version of Table 1 will be available online.
A portion is shown here to illustrate the table format and contents.
The \texttt{flag} column provides an auxiliary RGB-like/SSG-like classification from an independent analysis that will be presented in a forthcoming paper.
This flag is provided solely to identify sources for which the RGB-based distance calibration may be less suitable.
}
\end{deluxetable*}

\begin{figure}
    \centering
    \includegraphics[width=1\linewidth]{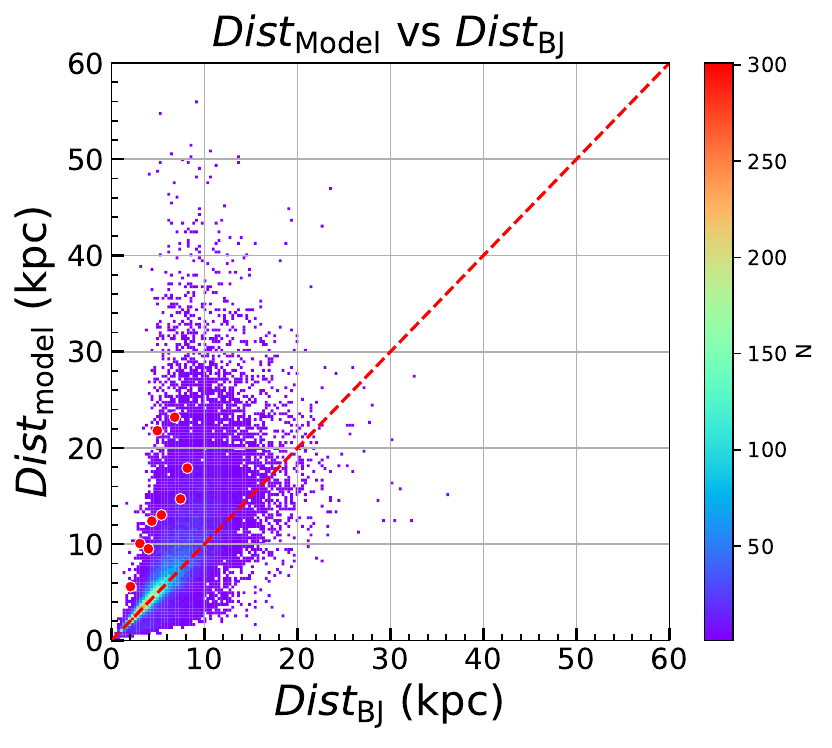}
    \caption{Comparison of $Dist_{\rm model}$ and $Dist_{\rm BJ}$, color-coded by number density. The red line marks the one-to-one relation. 
    Red dots mark the surviving HVS candidates.}
    \label{fig:dist_compare}
\end{figure}

\begin{figure*}
    \centering
    \includegraphics[width=1\linewidth]{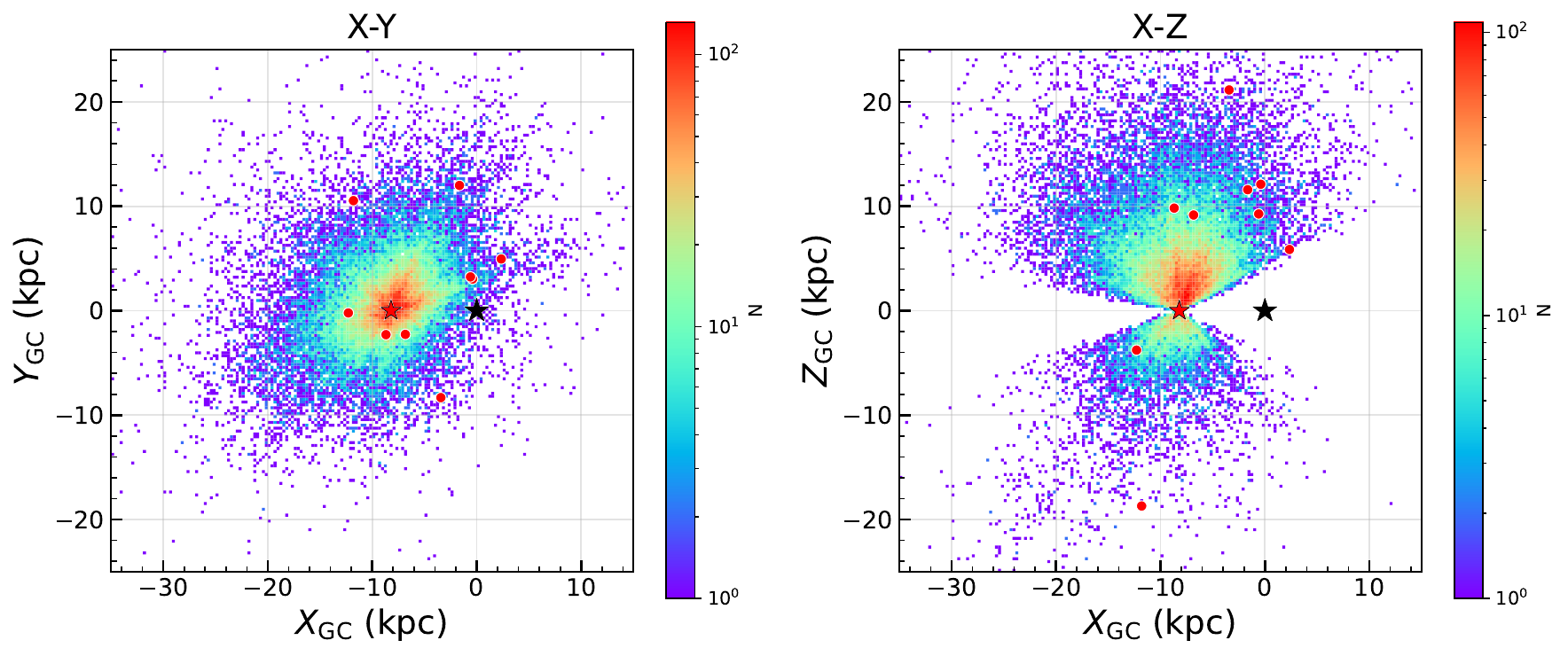}
    \caption{Galactocentric spatial distribution of the distance-catalog sample in the $X$--$Y$ (left) and $X$--$Z$ (right) planes, color-coded by number density. Red circles mark the nine surviving HVS candidates. The black star denotes the Galactic Center, and the red star marks the present position of the Sun.}
    \label{fig:space}
\end{figure*}

Figure~\ref{fig:dist_compare} compares $Dist_{\rm model}$ and $Dist_{\rm BJ}$. Within $\sim$10\,kpc, the two estimates are broadly consistent, whereas at larger distances $Dist_{\rm model}$ tends to exceed $Dist_{\rm BJ}$, consistent with the increasing influence of the Bayesian prior in \citet{bailer2021} at large distances (which can pull distant halo stars toward smaller inferred distances). A subset of stars instead shows $Dist_{\rm model}\ll Dist_{\rm BJ}$; inspection of their HR-diagram locations and spectra suggests that many are horizontal-branch and red clump (RC) stars.
The spatial distribution of the final sample is shown in Figure~\ref{fig:space}.

We further evaluate the model using six globular clusters spanning a range of metallicities. The cluster sample is drawn from \citet{Baumgardt2021}\footnote{\url{https://people.smp.uq.edu.au/HolgerBaumgardt/globular/}}. 
To construct a clean comparison set, we retain only sources with $p_{\rm single}=1$ and require $\mathrm{RUWE}\leq 1.2$ to minimize contamination from unresolved binaries.
Here, $p_{\rm single}=1$ indicates stars classified as single-star members in the \citet{Baumgardt2021}.
Cluster metallicities and distance are adopted from the Harris catalog \citep[2010 edition]{harris1996}\footnote{\url{https://physics.mcmaster.ca/~harris/mwgc.dat}}.

\begin{figure*}
    \centering
    \includegraphics[width=1\linewidth]{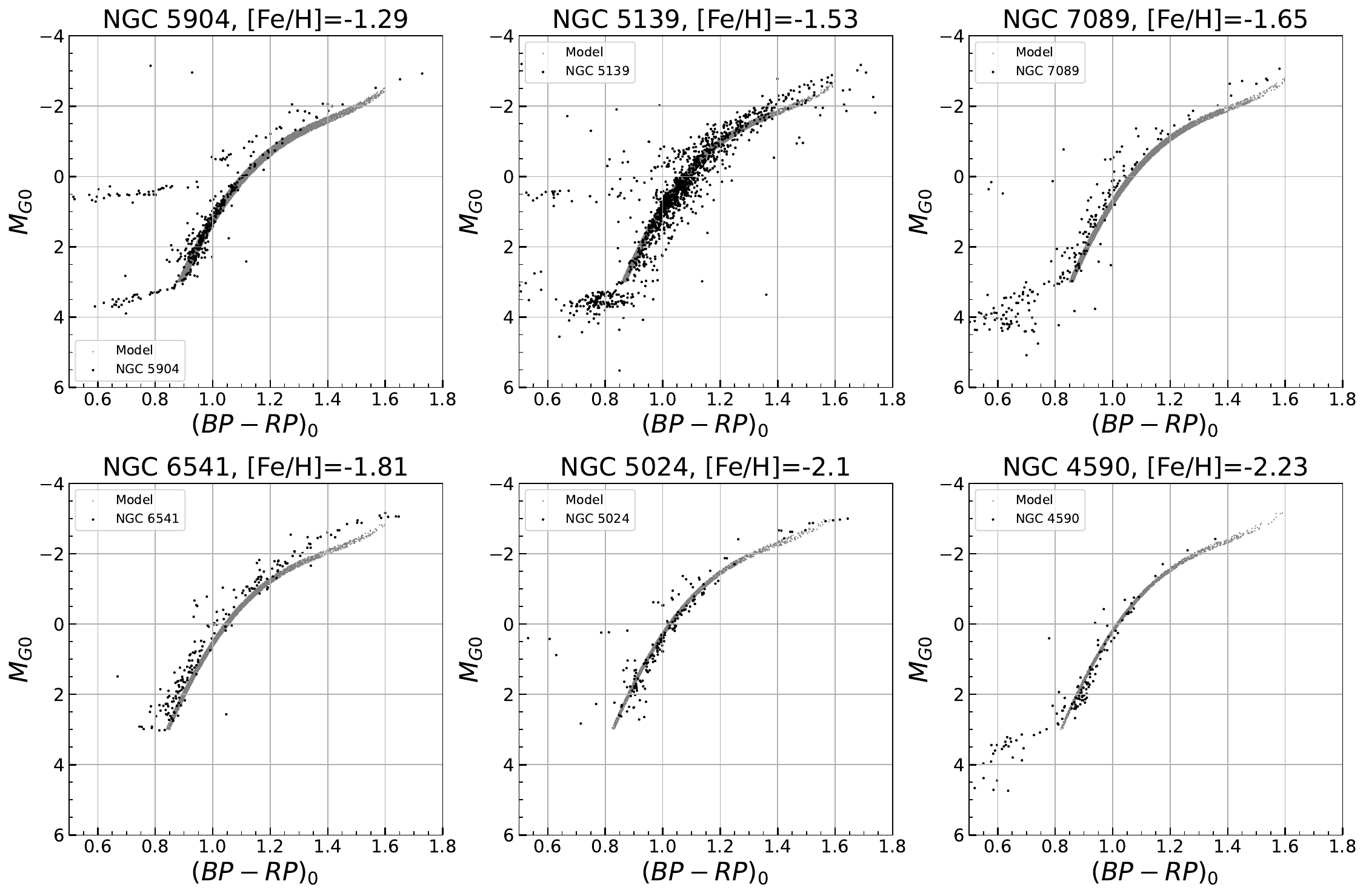}
    \caption{Comparison of the model-predicted RGB sequence with the observed HR diagrams of six globular clusters spanning a range of metallicities. Gray points show the model, while black points denote cluster stars from \citet{Baumgardt2021} after applying the quality cuts described in the text.}
    \label{fig:GC}
\end{figure*}

For each cluster, we select stars from our distance catalog with metallicities within $\pm 0.1$\,dex of the cluster value and compare their HR diagrams with the cluster sequence (Figure~\ref{fig:GC}). Overall, the model reproduces the globular-cluster sequences well. A systematic offset appears for very metal-poor (VMP, [Fe/H]$\le-2$; \citealt{beers2005}) red stars, for which the model tends to underestimate distances, likely due to sparse training coverage in that region of parameter space.

A caveat is that our calibration is constructed for normal metal-poor giants and thus need not apply to other stellar types in the parent sample. Assigning distances to non-giant contaminants with a giant-star relation can introduce systematic biases. For example, horizontal-branch and RC stars are typically assigned distances that are too small, leading to underestimated space velocities. Conversely, intrinsically fainter objects (e.g., sub-subgiant-like systems) can have their distances overestimated when interpreted as normal giants. Sub-subgiants occupy unusual HR-diagram locations---redward of the main sequence and fainter than normal subgiants or giants---and are thought to arise from non-standard evolutionary pathways, often involving binary interaction or magnetic activity (\citealt{geller2017}). Similar issues can occur for chromospherically active binaries, including RS~CVn-like systems. Using Gaia EDR3 photometry and parallaxes, \citet{leiner2022} showed that RS~CVn systems span a broad range of positions in the color--magnitude diagram, with many lying below the normal subgiant branch and redward of the RGB, indicating that they do not necessarily follow standard giant-branch relations. For this reason, after identifying high-velocity candidates, we inspect their spectra and stellar properties to remove likely contaminants.

\section{HVS candidates}\label{hvs}

Using Gaia DR3 positions and proper motions, LAMOST DR13 radial velocities, and the distances from our calibrated relation, we transform all 41{,}331 stars into the Galactocentric frame. We adopt the same Solar position $(-8.249,0,0)$\,kpc and Solar peculiar motion $(11.1,12.24,7.25)$\,\kms\ as in \citet{hong2024} and \citet{xu2025}. We then compute Galactocentric positions and total velocities and evaluate bound/unbound status with \texttt{AGAMA} \citep{Vasiliev2019}, adopting the \citet{McMillan2017} Galactic potential.

\begin{figure}
    \centering
    \includegraphics[width=1\linewidth]{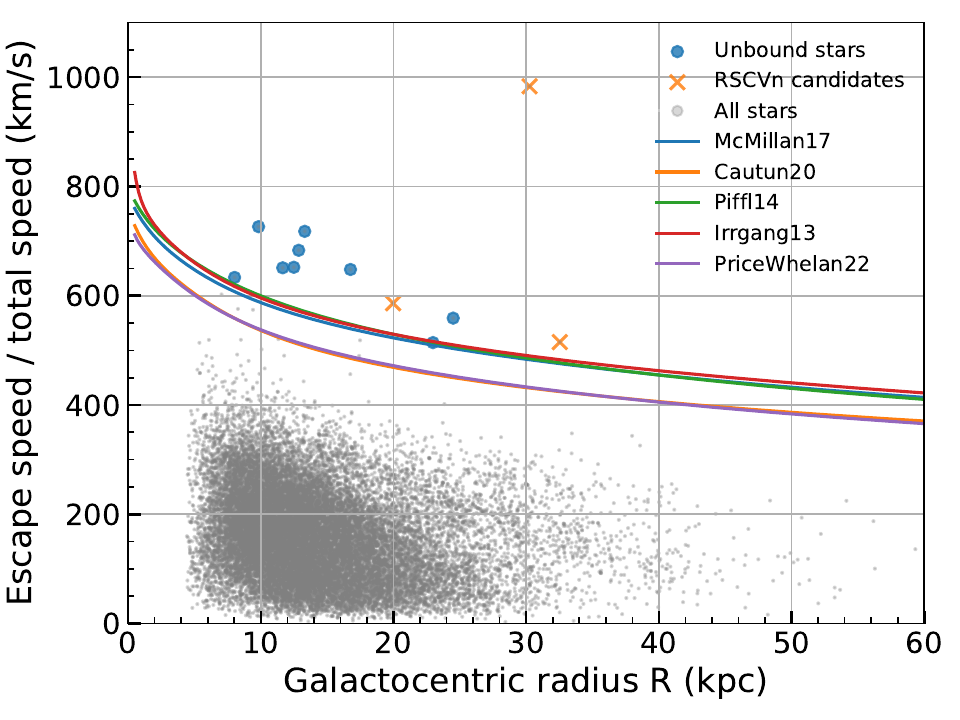}
    \caption{Total Galactocentric velocity as a function of Galactocentric radius for the distance sample. Gray points show all stars. Blue circles indicate the initial unbound candidates, and orange crosses mark sources later identified as RS~CVn candidates and removed from the final HVS sample. One orange cross lies outside the plotted velocity range at $(R_{\rm GC},v_{\rm GC})=(12.91\,\mathrm{kpc},1851.38\,\mathrm{km\,s^{-1}})$ and is not shown. Solid curves show the escape-speed profiles for five Galactic potentials (McMillan17, Cautun20, Piffl14, Irrgang13, and Price-Whelan22). 
    ALL candidates lie well above the corresponding escape-speed curves.}
    \label{fig:vel_R}
\end{figure}

Under the \citet{McMillan2017} potential, 13 sources have total velocities exceeding the local escape speed. 
Adopting the same Solar position and peculiar motion,
these sources also remain formally unbound in several alternative potentials, including \citet{irrgang2013}, \citet{piffl2014}, \citet{cautun2020}, and the \texttt{gala} MilkyWayPotential2022 (Price-Whelan et~al.\ 2022; see also \citealt{palladino2014}). Figure~\ref{fig:vel_R} shows total velocity versus Galactocentric radius for the full sample, together with escape-speed curves for the different potentials. The 13 initial candidates are highlighted in blue, and crosses mark objects later rejected as non-RGB contaminants.

As discussed above, photometric distances are highly sensitive to stellar type, so non-standard evolved objects can contaminate the initial HVS candidate list. We cross-match the 13 initial candidates against SIMBAD and find that four are classified as RS~CVn binaries. 
Ca\,II H\&K and H$_\alpha$ emission are classic signatures of enhanced chromospheric activity in RS~CVn binaries, in which tidal locking enforces rapid rotation, strengthening the magnetic dynamo and heating the chromosphere.
Their LAMOST spectra show clear chromospheric-activity signatures, including Ca\,II H\&K and H$_\alpha$ emission, indicating that a standard RGB calibration is unlikely to be reliable; we therefore exclude these four objects from the final HVS sample.

\begin{figure*}
    \centering
    \includegraphics[width=1\linewidth]{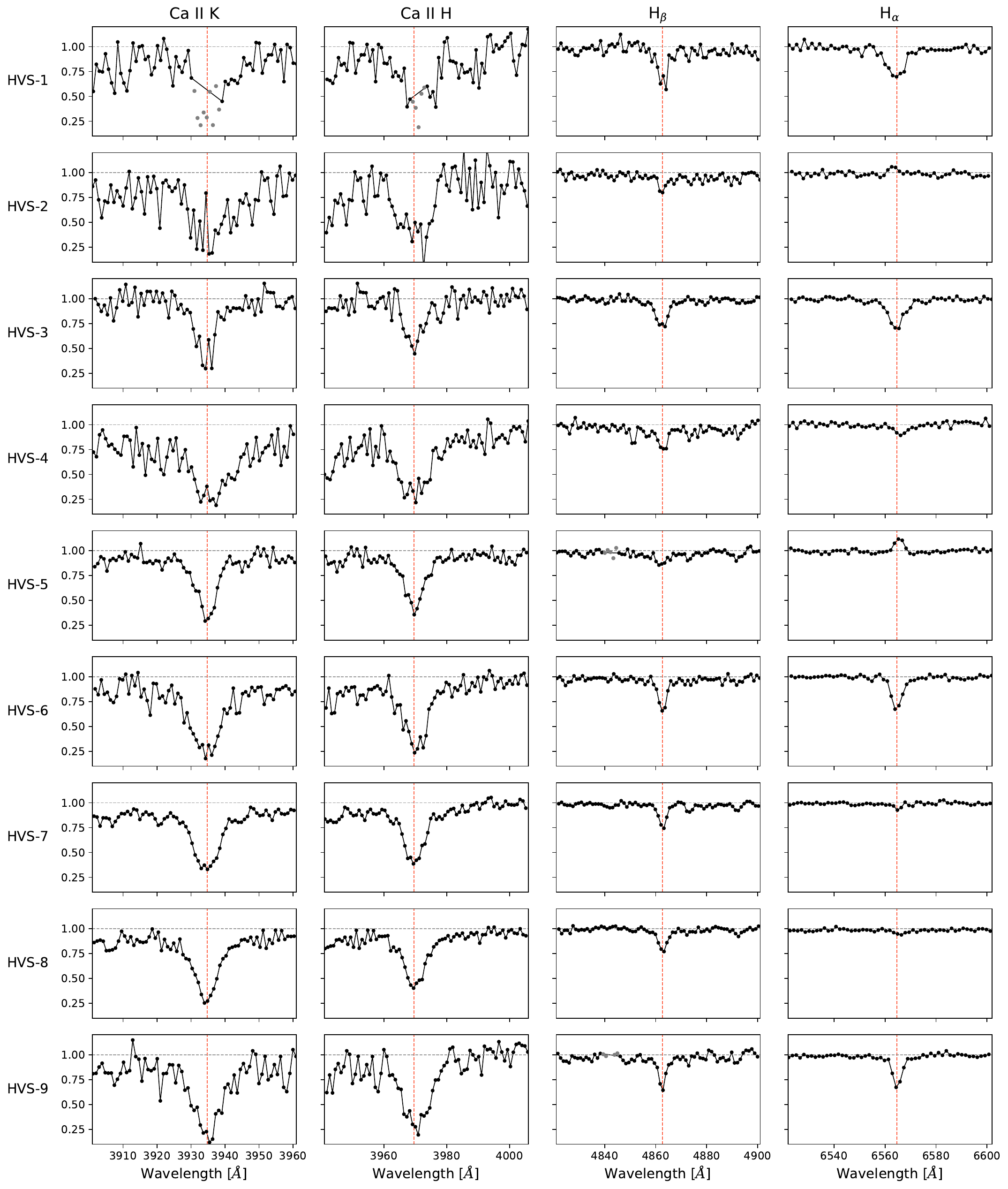}
    \caption{Normalized LAMOST spectra of the nine surviving HVS candidates in the Ca\,II K, Ca\,II H, H$_\beta$, and H$_\alpha$ windows. The spectra are shifted to the rest frame using the radial velocities in Table~\ref{tab:hvs_basic}. Vertical dashed red lines mark the laboratory wavelengths of the corresponding transitions, and horizontal dashed gray lines indicate the normalized continuum level. Black points and connecting lines show the spectra, while gray points denote problematic pixels flagged by the pipeline masks or inverse-variance filtering.}
    \label{fig:spec_activity}
\end{figure*}

For the remaining nine sources, we inspect activity-sensitive spectral windows (Ca\,II H\&K, H$_\beta$, and H$_\alpha$; Figure~\ref{fig:spec_activity}). Two show H$_\alpha$ emission (HVS-2 and HVS-5), two show partially filled-in H$_\alpha$ (HVS-7 and HVS-8), and five show normal H$_\alpha$ absorption. Because the sources with H$_\alpha$ emission or filling-in do not exhibit comparably strong Ca\,II H\&K emission, we do not reject them on this basis alone. The basic properties of the nine surviving candidates are listed in Table~\ref{tab:hvs_basic}.
Uncertainties in $V_{\rm tot}$ are estimated via Monte Carlo sampling, propagating the asymmetric photometric-distance uncertainties implied by the intrinsic model scatter in $M_G$ (9.7\%), Gaia proper-motion uncertainties (including the $\mu_{\alpha*}$--$\mu_\delta$ covariance), and LAMOST radial-velocity uncertainties.

\begin{deluxetable*}{cccccccccccc}
\tabletypesize{\scriptsize}
\tablecaption{Properties of the nine surviving HVS candidates.\label{tab:hvs_basic}}
\tablewidth{0pt}
\tablehead{
\colhead{ID} &
\colhead{Gaia DR3 ID} &
\colhead{RA} &
\colhead{Dec} &
\colhead{$\varpi$} &
\colhead{$\varpi/\sigma_\varpi$} &
\colhead{$\mu_{\alpha *}$} &
\colhead{$\mu_{\delta}$} &
\colhead{$v_{\rm rad}$} &
\colhead{[Fe/H]} &
\colhead{$(BP-RP)_0$} &
\colhead{Remark}
}
\startdata
HVS-1 & 3264514486031687552 & 52.17897  & 0.73961  & 0.461 & 4.51 & 30.32  & $-$0.24  & 206.58  & $-$1.74 & 0.83 & Normal RGB\\
HVS-2 & 2548501931225760128 & 5.59134   & 2.73374  & 0.112 & 1.41 & 5.79   & $-$4.72  & $-$53.63  & $-$1.43 & 1.06 & $H_\alpha$ emission\\
HVS-3 & 3949683746815244160 & 182.80209 & 17.32701 & 0.274 & 2.52 & $-$14.54 & $-$13.64 & 23.63   & $-$2.37 & 0.85 & Normal RGB\\
HVS-4 & 3703665236962337152 & 191.43563 & 2.81250  & 0.016 & 0.21 & 1.41   & $-$4.91  & $-$143.03 & $-$1.77 & 1.05 & Normal RGB\\
HVS-5 & 3927394962493384576 & 192.19389 & 11.05466 & 0.190 & 3.17 & $-$1.67  & $-$18.59 & $-$34.57  & $-$2.41 & 0.90 & $H_\alpha$ emission\\
HVS-6 & 4392803631644275712 & 255.60530 & 5.62728  & 0.185 & 2.87 & $-$0.51  & $-$12.90 & $-$113.17 & $-$1.63 & 0.99 & Normal RGB\\
HVS-7 & 1352334469635055488 & 251.42104 & 38.27802 & 0.082 & 2.65 & $-$0.89  & $-$8.68  & $-$154.89 & $-$1.88 & 1.08 & $H_\alpha$ filled-in\\
HVS-8 & 1184128157160434560 & 228.15515 & 15.90453 & 0.080 & 1.97 & 0.66   & $-$11.29 & $-$88.84  & $-$2.09 & 1.01 & $H_\alpha$ filled-in\\
HVS-9 & 1194066986361119232 & 235.28224 & 13.89201 & 0.203 & 3.24 & 0.14   & $-$14.79 & 5.47    & $-$1.67 & 0.98 & Normal RGB\\[3pt]
\hline
\hline\\[-5pt]
ID & $Dist_{\rm model}$ & $X$ & $Y$ & $Z$ & $V_X$ & $V_Y$ & $V_Z$ & $R_{\rm GC}$ & $V_{\rm tot}$ & $V_{\rm esc}$ & Likely origin\\[5pt]
\hline
HVS-1 & $5.6^{+0.6}_{-0.5}$  & $-$12.3 & $-$0.2 & $-$3.8  & $-$528.0 & $-$294.8 & 317.7  & 12.3 & $683.2^{+76.7}_{-68.3}$ & 562.1 & \\
HVS-2 & $21.8^{+2.2}_{-2.0}$ & $-$11.8 & 10.5   & $-$18.7 & $-$261.9 & $-$422.5 & $-$255.7 & 15.8 & $559.0^{+78.1}_{-71.0}$ & 501.9 & Disk\\
HVS-3 & $10.1^{+1.0}_{-0.9}$ & $-$8.7  & $-$2.3 & 9.8     & $-$272.1 & $-$636.4 & $-$190.4 & 9.0  & $717.8^{+96.4}_{-87.0}$ & 556.2 & \\
HVS-4 & $23.2^{+2.4}_{-2.1}$ & $-$3.4  & $-$8.3 & 21.1    & 381.8    & $-$29.4  & $-$343.2 & 9.0  & $514.2^{+48.7}_{-43.5}$ & 507.5 & \\
HVS-5 & $9.5^{+1.0}_{-0.9}$  & $-$6.8  & $-$2.3 & 9.2     & 381.2    & $-$459.3 & $-$260.0 & 7.2  & $651.1^{+83.1}_{-75.2}$ & 567.0 & \\
HVS-6 & $13.0^{+1.3}_{-1.2}$ & 2.4     & 5.0    & 5.9     & 367.2    & $-$379.8 & $-$349.6 & 5.5  & $633.5^{+78.9}_{-72.1}$ & 597.3 & Sagittarius\\
HVS-7 & $17.9^{+1.8}_{-1.7}$ & $-$1.7  & 12.0   & 11.6    & 606.1    & $-$211.0 & $-$89.5  & 12.1 & $648.0^{+73.4}_{-66.8}$ & 536.5 & \\
HVS-8 & $14.7^{+1.5}_{-1.4}$ & $-$0.4  & 3.0    & 12.1    & 505.5    & $-$302.5 & $-$279.5 & 3.1  & $652.1^{+77.0}_{-69.2}$ & 560.3 & Sagittarius\\
HVS-9 & $12.4^{+1.3}_{-1.1}$ & $-$0.6  & 3.2    & 9.3     & 584.8    & $-$355.8 & $-$242.8 & 3.3  & $726.3^{+85.9}_{-77.2}$ & 579.6 & Sagittarius\\
\enddata
\tablenotetext{}{The upper part lists observed properties of the nine surviving HVS candidates: R.A. and Dec (deg), parallax $\varpi$ (mas), proper motions $\mu_{\alpha *}$ and $\mu_\delta$ (mas\,yr$^{-1}$), radial velocity $v_{\rm rad}$ (\kms), metallicity $\mathrm{[Fe/H]}$ (dex), and intrinsic color $(BP-RP)_0$ (mag). The lower part lists derived quantities: $Dist_{\rm model}$, $X$, $Y$, $Z$, and $R_{\rm GC}$ (kpc), and $V_X$, $V_Y$, $V_Z$, $V_{\rm tot}$, and $V_{\rm esc}$ (\kms). The final column reports the probable origin where available.}
\end{deluxetable*}

As Figure~\ref{fig:dist_compare} shows, all nine candidates satisfy $\mathrm{Dist}_{\rm model} > \mathrm{Dist}_{\rm BJ}$. This follows naturally: for fixed proper motions and radial velocities, adopting $\mathrm{Dist}_{\rm model}>\mathrm{Dist}_{\rm BJ}$ increases the inferred tangential velocity, making an unbound classification more likely. Because sources with reliable Gaia astrometry rarely yield robust HVS candidates \citep{Marchetti2022}, the lack of high-significance parallaxes in our sample is not surprising. The candidates indeed have low-significance Gaia parallaxes, with $\varpi/\sigma_{\varpi}$ ranging from 0.21 to 4.51 (and below 3 for most sources; Table~\ref{tab:hvs_basic}). Consequently, the Bailer--Jones distances are only weakly constrained by parallax and are more sensitive to the adopted Galactic prior, which can bias $\mathrm{Dist}_{\rm BJ}$ toward smaller values.

\begin{figure*}
    \centering
    \includegraphics[width=1\linewidth]{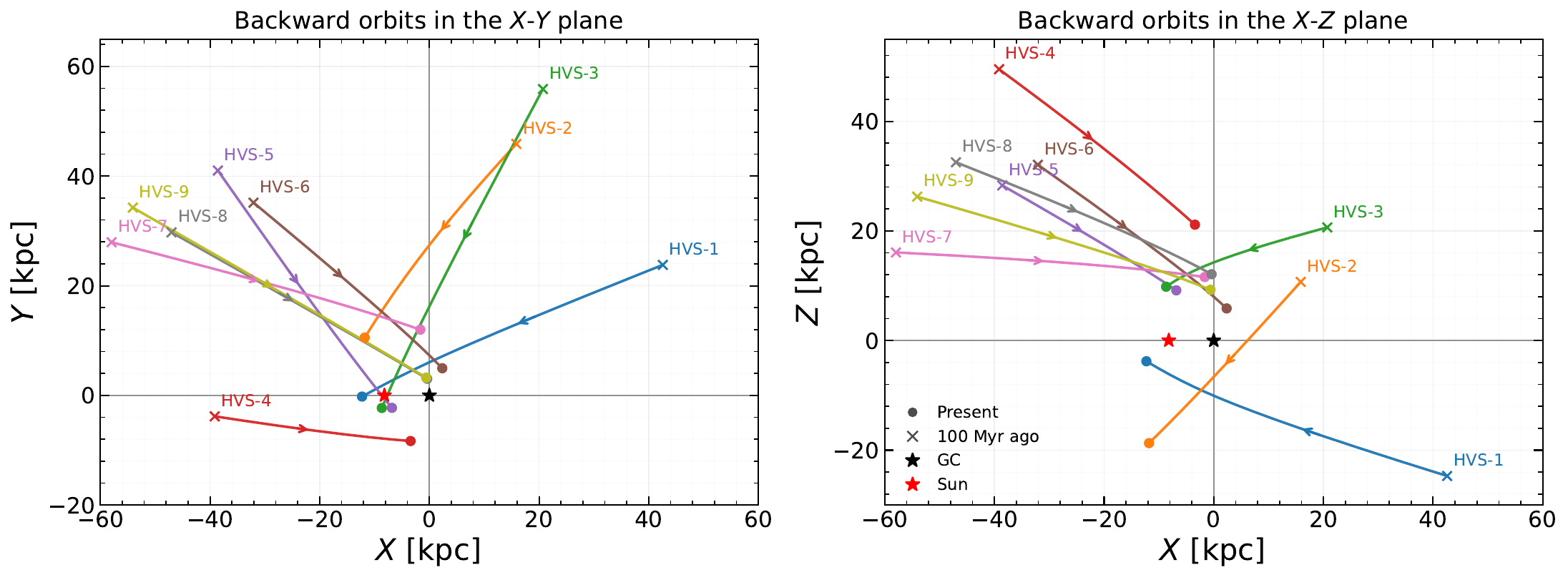}
    \caption{Backward-integrated orbits of the nine surviving HVS candidates in the $X$--$Y$ (left) and $X$--$Z$ (right) planes, integrated for 100\,Myr in the \citet{McMillan2017} Galactic potential. Colored circles mark present-day positions, crosses mark positions 100\,Myr ago, and arrows indicate the direction of motion along the backward-integrated trajectories. The black star denotes the Galactic Center, and the red star marks the present position of the Sun.}
    \label{fig:orbit}
\end{figure*}

We then integrate the orbits of the nine candidates backward for 100\,Myr in the \citet{McMillan2017} potential and show their trajectories in the $X$--$Y$ and $X$--$Z$ planes in Figure~\ref{fig:orbit}. 
Only one candidate plausibly originates in the Galactic disk; the remainder are more consistent with halo-like origins, as expected given our focus on metal-poor giants.

Interestingly, three candidates trace back toward a similar direction on the sky. Their present-day positions overlap the spatial extent of the Sagittarius stream mapped by \citet{ramos2022}, and their orbital angular momenta and metallicities are consistent with typical Sagittarius-stream stars. Taken together, these properties indicate that the three objects most likely originated in the Sagittarius system.

The lack of a clear Galactic-center origin argues against a classical Hills-mechanism interpretation for the final sample. Instead, the candidates appear to trace a mixture of origins. The disk-compatible object is consistent with a runaway-star scenario, such as binary-supernova ejection or dynamical ejection from a dense stellar environment in the disk. The three Sagittarius-like candidates are more naturally interpreted as possible Sagittarius-associated debris, rather than Galactic-center ejecta. The remaining halo-like objects cannot be tied to a unique progenitor with the present data, but may be associated with accreted halo substructure rather than a single common ejection site.

Our final sample of nine candidates appears to be entirely new, with no obvious object-by-object overlap with previously reported HVS or extreme-velocity-star samples. This provides a concrete illustration of the sensitivity of dynamical classification to the adopted distance: the transition from Gaia DR2 to later Gaia releases can substantially revise earlier HVS identifications. Indeed, many DR2-era candidates were re-assessed once improved astrometry became available, and several proposed late-type HVSs were subsequently shown to be more likely bound to the Milky Way or otherwise inconsistent with a Galactic-center origin (e.g., \citealt{boubert2018,Marchetti2019,marchetti2021}). Thus, later Gaia releases did not simply expand the candidate pool, but also prompted major re-evaluation of previously reported objects.

The limited overlap with Gaia DR3-era samples is most naturally explained by differences in sample definitions and distance methodologies. The Gaia DR3 6D search of \citet{Marchetti2022} was intentionally restricted to stars with precise parallaxes, enabling purely geometric distance estimates from Gaia astrometry; this conservative choice disfavors the distant metal-poor giants that motivate our photometric calibration. The samples of \citet{luna2024} and \citet{liao2023} are dominated by RC stars, with distances and tangential velocities derived under the RC standard-candle assumption, whereas our selection largely excludes RC stars. Similarly, the sample of \citet{li2023} includes many dwarfs and hot stars outside the domain of our metal-poor giant calibration, which we remove by construction. The lack of common sources therefore likely reflects differences in distance scales and targeted stellar populations, rather than any inconsistency in the overall kinematic picture.

\section{Conclusions}\label{summary}

In this work, we search for HVS candidates among metal-poor giants by combining LAMOST DR13 spectroscopy with Gaia DR3 astrometry. We construct a metallicity-dependent color--absolute-magnitude relation for normal metal-poor giants using a calibration sample of 6298 stars, with reference absolute magnitudes from the Bailer-Jones distance estimates. After removing horizontal-branch stars and subgiants, the resulting relation reproduces the reference magnitudes with a scatter of 0.24\,mag, corresponding to an intrinsic distance uncertainty of $\sim$9.7\%.

Applying this calibration to the full parent sample yields photometric distances for 41{,}331 metal-poor giants, spanning distances from 0.16 to 100\,kpc, with 90\% of the sample within 16\,kpc.
Relative to the Bailer-Jones distances, the two scales agree well at $\lesssim 10$\,kpc but diverge systematically at larger distances. Tests on six globular clusters show that the model generally reproduces observed RGB sequences, although it tends to underestimate distances for the reddest and most metal-poor stars, where the training coverage is sparse.

Combining these distances with Gaia DR3 proper motions and LAMOST DR13 radial velocities, we identify 13 initially unbound candidates under the McMillan17 Galactic potential. Spectral inspection and SIMBAD cross-matching indicate that four are likely RS~CVn binaries or other chromospherically active systems for which a normal-giant calibration is inappropriate; removing these leaves a final sample of nine metal-poor HVS candidates. Backward orbit integrations over 100\,Myr suggest that one candidate is most consistent with a disk origin and three are plausibly associated with the Sagittarius stream, while the remainder are halo-like high-velocity objects with uncertain origins. Although these nine stars do not necessarily constitute a single, uniformly characterized population, they demonstrate that a giant-based distance calibration can recover a set of promising high-velocity candidates that is largely distinct from those found in Gaia-parallax-dominated searches.

Our final sample shows no overlap with previously reported HVS or extreme-velocity-star samples, which is naturally explained by differences in distance methodology and sample definition. Many Gaia DR2-era late-type candidates were re-evaluated with improved astrometry and subsequently found more likely to be bound to the Milky Way. For Gaia DR3-era studies, limited overlap is likewise expected: some searches focused on RC stars, while others included large numbers of dwarfs, both outside the primary domain of our metal-poor giant calibration.

Looking ahead, the combination of Gaia with large spectroscopic surveys will continue to improve HVS searches across a wider range of stellar populations. Metal-poor giants are particularly valuable tracers because their high luminosities enable detections at large heliocentric distances, where extreme-velocity candidates are most readily identified. For this regime, LAMOST is especially important: the LAMOST~III target-selection strategy has strengthened the sampling of metal-poor and VMP stars, expanding the discovery space for HVSs (S. Xu et~al., in prep.).
Complementary spectroscopy from DESI \citep{DESI2016} and 4MOST \citep{deJong2019}, together with improved astrometry and an expanded radial-velocity sample from Gaia DR4, will enable re-assessment of marginal candidates, a more complete census of metal-poor HVSs, and tighter constraints on the relative roles of different origin channels.

\begin{acknowledgments}
This work is supported by the National Key Basic R\&D Program of China via 2024YFA1611901 and 2024YFA1611601, the National Natural Science Foundation of China through the projects NSFC 12222301, 12173007, 124B2055, 12573032, and 12403024.  
Guoshoujing Telescope (the Large Sky Area Multi-Object Fiber Spectroscopic Telescope, LAMOST) is a National Major Scientific Project built by the Chinese Academy of Sciences. Funding for the project has been provided by the National Development and Reform Commission. LAMOST is operated and managed by the National Astronomical Observatories, Chinese Academy of Sciences. 
This work has made use of data from the European Space Agency (ESA) mission {\it Gaia} (\url{https://www.cosmos.esa.int/gaia}), processed by the Gaia Data Processing and Analysis Consortium (DPAC, \url{https:// www.cosmos.esa.int/web/gaia/dpac/ consortium}). Funding for the DPAC has been provided by national institutions, in particular, the institutions participating in the Gaia Multilateral Agreement.
\end{acknowledgments}

\bibliographystyle{aasjournal}
\bibliography{HVS}

\end{CJK*}
\end{document}